\documentclass[10pt,conference]{IEEEtran}

\IEEEoverridecommandlockouts

\usepackage{amsmath,amssymb,amsfonts}
\usepackage{algorithmic}
\usepackage{graphicx}
\usepackage{textcomp}
\usepackage{xcolor}
\usepackage{ragged2e}
\usepackage{balance}
\usepackage{colortbl}
\usepackage{multirow}
\usepackage{balance}
\usepackage{color}

\usepackage{CJKutf8}

\usepackage[colorlinks,
            linkcolor=blue,
            anchorcolor=blue,
            citecolor=blue]{hyperref}

\begin{document}

\title{Multi-Programming-Language Commits in OSS: \\An Empirical Study on Apache Projects}

\author{
    \IEEEauthorblockN{Zengyang Li\IEEEauthorrefmark{2}\thanks{This work was partially funded by the National Natural Science Foundation of China under Grant Nos. 61702377 and 62002129.}, Xiaoxiao Qi\IEEEauthorrefmark{2}, Qinyi Yu\IEEEauthorrefmark{2}, Peng Liang\IEEEauthorrefmark{3}\IEEEauthorrefmark{1}\thanks{\IEEEauthorrefmark{1} Corresponding author}, Ran Mo\IEEEauthorrefmark{2}, Chen Yang\IEEEauthorrefmark{4}}
    \IEEEauthorblockA{
    \IEEEauthorrefmark{2}School of Computer Science \& Hubei Provincial Key Laboratory of Artificial Intelligence and Smart Learning, \\Central China Normal University, Wuhan, China \\
    \IEEEauthorrefmark{3}School of Computer Science, Wuhan University, Wuhan, China\\
    \IEEEauthorrefmark{4}IBO Technology (Shenzhen) Co., Ltd., Shenzhen, China\\
    \{zengyangli, moran\}@ccnu.edu.cn, \{qixiaoxiao, qinyiyu\}@mails.ccnu.edu.cn, liangp@whu.edu.cn, c.yang@ibotech.com.cn}
}

\maketitle

\begin{abstract}
Modern software systems, such as Spark, are usually written in multiple programming languages (PLs). Besides benefiting from code reuse, such systems can also take advantages of specific PLs to implement certain features, to meet various quality needs, and to improve development efficiency. In this context, a change to such systems may need to modify source files written in different PLs. We define a multi-programming-language commit (MPLC) in a version control system (e.g., Git) as a commit that involves modified source files written in two or more PLs. To our knowledge, the phenomenon of MPLCs in software development has not been explored yet. In light of the potential impact of MPLCs on development difficulty and software quality, we performed an empirical study to understand the state of MPLCs, their change complexity, as well as their impact on open time of issues and bug proneness of source files in real-life software projects. By exploring the MPLCs in 20 non-trivial Apache projects with 205,994 commits, we obtained the following findings: (1) 9\% of the commits from all the projects are MPLCs, and the proportion of MPLCs in 80\% of the projects goes to a relatively stable level; (2) more than 90\% of the MPLCs from all the projects involve source files written in two PLs; (3) the change complexity of MPLCs is significantly higher than that of non-MPLCs in all projects; (4) issues fixed in MPLCs take significantly longer to be resolved than issues fixed in non-MPLCs in 80\% of the projects; and (5) source files that have been modified in MPLCs tend to be more bug-prone than source files that have never been modified in MPLCs. These findings provide practitioners with useful insights on the architecture design and quality management of software systems written in multiple PLs.
\end{abstract}


\section{Introduction}
\label{chap:intro}
Modern software systems, such as Apache Spark and Ambari, are usually written in multiple programming languages (PLs). One of the main reasons for adopting multiple PLs is to reuse existing code with required functionalities \cite{GrAbJa2020}. Another main reason is to take advantages of specific PLs to implement certain features, to meet various software quality needs, and to improve software development efficiency \cite{GrAbJa2020, RaPoFiDe2014, MaBa2015,KoWiLo2016,Ma2017,MaKiLe2017,AbGrKh2019}. Nowadays, multi-programming-language (MPL) software development is increasingly prevalent with the technology advances \cite{KoLiWo2006,KoWiLo2016,AbRaOpKh2021}. 

However, despite great benefits of MPL systems, they are also facing various challenges. For example, static code analysis is much more difficult in MPL systems than mono-language systems since multiple PLs and cross-language communication mechanisms need to be analyzed simultaneously \cite{ShMiAb2019, kaIsIz2019}.

In an MPL system, there inevitably exist a certain proportion of code changes that need to modify source files written in different PLs. Intuitively, a code change in which the modified source files are written in multiple PLs is likely to modify more than one component in a software system, and thus the complexity of such a change is relatively high. As a result, such a code change may need more effort and time to understand and analyze the impact on the parts affected by the modified source files. We define an MPL commit (MPLC) in a version control system (e.g., Git) as a commit that involves modified source files written in two or more PLs. 

Although there are a few studies that investigated the quality of MPL systems \cite{RaPoFiDe2014, KoWiLo2016, GrEgAd2020}, these studies took a project or pull request as an analysis unit, which is at a relatively high level and may not provide specific advice for software development practice. In contrast, a commit is in a finer granularity than a project or pull request, and developers deal with code changes in commits in daily development practices. Hence, we suggest to investigate MPL software development from the perspective of commits (i.e., MPLCs). To our knowledge, the phenomenon of MPLCs in software development has not been explored yet. Considering the potential impact of MPLCs on development difficulty (e.g., cross-language change impact analysis \cite{ShMiAb2019}) and software quality (e.g., bug introduction \cite{KoWiLo2016, BeHoMaViVi2019}), we conducted an empirical study to understand the state of MPLCs, their change complexity, as well as their impact on open time of issues and bug proneness of source files in real-life software projects, which provides a foundation for improving the practices of MPL software development.

The main contributions are summarized as follows:
 
 \begin{itemize}
  \item This work is a first attempt to explore the phenomenon of MPLCs in real-world settings.
  \item The state of MPLCs (including the proportion of MPLCs and the number of PLs used in MPLCs) in MPL software systems is explored. 
  \item The change complexity of MPLCs, open time of issues fixed in MPLCs, and bug proneness of source files modified in MPLCs in MPL systems are studied in depth.
\end{itemize}

The remaining of this paper is organized as follows. Section \ref{chap:relat} presents the related work; Section \ref{chap:case} describes the design of the empirical study; Section \ref{chap:study} presents the results of the study; Section \ref{chap:discus} discusses the study results; Section \ref{chap:threats} identifies the threats to validity of the results; and Section \ref{conclusions} concludes this work with future research directions.

\section{Related Work}\label{RelatedWork}
\label{chap:relat}

{To the best of our knowledge, there has not been work on MPLCs. Thus, the related work presented here is not directly relevant to MPLCs, but related to the research on MPL software systems in general. The related work is presented in two aspects, including the phenomenon of MPL software systems and quality of MPL software systems.}

\subsection{Phenomenon of MPL Software Systems }\label{RelatedWork_A}
Mayer and Bauer studied the phenomenon of multi-language programming using data mining technologies on 1,150 open source software (OSS) projects gathered from GitHub \cite{MaBa2015}. They used the Poisson regression model to explore the relationship between the number of PLs of the project and the size, age, number of contributors, and number of commits. They found that each project uses an average of 5 PLs with a clearly dominant PL; the median number of general-purpose PLs and domain-specific PLs is 2 and 2, respectively. The results also confirmed that the use of multiple PLs is very common in OSS projects. The focus of our work is different in that we investigated MPL software systems from the perspective of commits, while the work of Mayer and Bauer paid more attention at the level of project; in addition, we went deeper and looked into the bug proneness and change complexity of source files modified in MPLCs as well.

\subsection{Quality of MPL Software Systems}\label{RelatedWork_B}
In 2011, Bhattacharya and Neamtiu investigated the impact of C and C++ on software maintainability and code quality of four large OSS projects \cite{BhNe2011}. They found that C++ code has higher internal quality than C code and C++ code is less prone to bugs than C code, but they could not confirm that C++ code needs less effort to maintain than C code. This work looked into the impact of specific PLs on code quality, while our work investigated the impact of MPLCs on the bug proneness of source files in terms of defect density.

In 2014, Ray \emph{et al}. studied the effect of PLs on software quality using a large dataset of 729 projects in 17 PLs gathered from GitHub \cite{RaPoFiDe2014}. They combined multiple regression modeling with visualization and text analytics to study the impact of language characteristics. They found that language design indeed has a significant but modest effect on software defects. In addition, they found that there is a small but significant correlation between language set and software defects. Specifically, they found that there are 11 PLs that have a relationship with software defects. In 2019, Berger \emph{et al}. \cite{BeHoMaViVi2019} carried out repeated experiments of the study of Ray \emph{et al}. \cite{RaPoFiDe2014} and reduced the number of defect-related languages down from 11 to only 4. These studies paid attention to the impact of language features on bug proneness of software systems. In contrast, our work is focused on the impact of MPLCs on the bug proneness of source files.

In 2016, Kochhar \emph{et al}. conducted a large-scale empirical investigation of the use of multiple PLs and the combination of certain PLs on bug proneness \cite{KoWiLo2016}. They analyzed a dataset comprised of 628 projects collected from GitHub, in 17 general-purpose PLs (e.g., Java and Python). They found that implementing a project with more PLs significantly increases bug proneness, especially on memory, concurrency, and algorithm bugs. The results also revealed that the use of specific PLs together is more bug-prone in an MPL setting. However, our work is focused on the development difficulty and software quality in a finer granularity of commits.

In 2019, Abidi \emph{et al}. identified six anti-patterns \cite{AbKhGu2019a} and twelve code smells \cite{AbKhGu2019b} in MPL software systems. Six anti-patterns were identified in OSS systems, including excessive inter-language communication, too much scattering, and so forth \cite{AbKhGu2019a}. Twelve code smells were proposed, including passing excessive objects, memory management mismatch, and so on \cite{AbKhGu2019b}. Abidi \emph{et al}. subsequently proposed an approach to detect aforementioned anti-patterns and code smells (both called design smells according to the authors) in MPL systems in which Java Native Interface (JNI) is used, and conducted an empirical study on the fault proneness of such MPL design smells in nine open source JNI projects \cite{AbRaOpKh2021}. They found that MPL design smells are prevalent in the selected projects and files with MPL design smells can often be more associated with bugs than files without these design smells, and that specific smells are more correlated to fault proneness than other smells. These design smells provide useful suggestions in practice to avoid design defects and implementation flaws in software development.

In 2019, Kargar \emph{et al}. proposed an approach to modularization of MPL applications \cite{kaIsIz2019}. The results show that the proposed approach can build a modularization close to human experts, which may be helpful in understanding MPL software systems. In 2020, the same authors proposed a method to improve the modularization quality of heterogeneous MPL software systems by unifying structural and semantic concepts \cite{kaIsIz2020}. Admittedly, architecture quality (e.g., modularity) of MPL software systems is worth further and deeper research. This study provided an important viewpoint for the research on MPL software systems. Based on the results of our work, we will examine the architecture quality of MPL software systems with a relatively high proportion of MPLCs.

In 2020, Grichi \emph{et al}. performed a case study on the impact of interlanguage dependencies in MPL systems \cite{GrAbJa2020}. They found that the risk of bug introduction gets higher when there are more interlanguage dependencies, while this risk remains constant for intralanguage dependencies; the percentage of bugs found in interlanguage dependencies is three times larger than the percentage of bugs identified in intralanguage dependencies.
Grichi \emph{et al}. also conducted a study on the impact of MPL development in machine learning frameworks \cite{GrEgAd2020}. They found that mono-language pull requests in machine learning frameworks are more bug-prone than traditional software systems. Their work investigated the impact of MPL code changes on software systems in a granularity of pull requests, while our work studied the bug proneness of MPL code changes on software systems in a finer granularity of commits. In addition, our work demonstrates a higher bug proneness of source files modified in MPLCs than source files modified in only non-MPLCs, which is a dramatic difference from the results obtained by the work of Grichi \emph{et al}. 

\section{Study Design}
\label{chap:case}
In order to investigate the state of MPLCs and their impact on development difficulty and software quality, we performed a case study on Apache OSS projects. The main reason for conducting a case study is that, through using OSS projects, and more specifically their commit records and issues, we can examine the phenomenon in its real-life context (i.e., OSS development), since both the commit records and issues cannot be monitored in isolation, and their environment cannot be controlled. In this section we describe the study, which was designed and reported following the guidelines proposed by Runeson and H{\"o}st \cite{RuHo2009}.
\subsection{Objective and Research Questions}\label{DesignRQ}
The goal of this study, described using the Goal-Question-Metric (GQM) approach \cite{Ba1992}, is: to \emph{analyze} commits and their involving source files and corresponding fixed issues \emph{for the purpose of} investigation \emph{with respect to} the state of MPLCs as well as their impact on development difficulty and software quality, \emph{from the point of view of} software developers \emph{in the context of} MPL OSS development.

Based on the abovementioned goal, we have formulated five research questions (RQs), which are classified into three categories and described as follows. 

~\\
\noindent \textbf{Category I}: State of MPLCs.

\noindent \textbf{RQ1}: What is the proportion of MPLCs over all commits of a project? 

\noindent \textbf{Rationale}: With this RQ, we investigate the frequency of MPLCs occurred in  software projects and how the proportion of MPLCs evolves, so as to get a basic understanding on the state of MPLCs in MPL software projects.

\noindent \textbf{RQ2}: How many programming languages are used in the modified source files of MPLCs?

\noindent \textbf{Rationale}: This RQ is focused on the number of PLs used in source files modified in MPLCs, which enables us to understand the tendency of the use of multiple PLs.

\noindent \textbf{Category II}: Impact of MPLCs on development difficulty.

\noindent \textbf{RQ3}: What is the code change complexity of MPLCs? Is there a difference on the code change complexity between MPLCs and non-MPLCs?

\noindent \textbf{Rationale}: To explore the development difficulty of MPLCs, we look into the code change complexity of MPLCs. Intuitively, the complexity of code changes in MPLCs may be different from that in non-MPLCs. With this RQ, we intend to calculate the complexity of code changes in MPLCs and further validate if the complexity of code changes in MPLCs is significantly higher than that in non-MPLCs. In addition, the complexity of code change in a commit can be measured by the number of lines of code, source files, and directories that are modified in the commit, and by the entropy \cite{Ha2009} of the modified files in the commit. These change complexity measures are adopted from \cite{LiLiLiMoLi2020}.

\noindent \textbf{RQ4}: Do the issues fixed in MPLCs tend to take longer to be resolved than issues fixed in non-MPLCs?

\noindent \textbf{Rationale}: With this RQ, we further investigate the time taken to resolve issues that were fixed in MPLCs and non-MPLCs. The time taken to resolve an issue can, to some extent, reflect the development difficulty of MPL software systems.

\noindent \textbf{Category III}: Impact of MPLCs on software quality.

\noindent \textbf{RQ5}: Are source files that have been modified in MPLCs more bug-prone than source files that have never been modified in MPLCs?

\noindent \textbf{Rationale}: MPLCs may influence the quality of software systems. With this RQ, we are concerned with the impact of MPLCs on software systems in terms of the likelihood of bugs.

\subsection{Cases and Unit Analysis}\label{CasesandUnitAnalysis}
According to \cite{RuHo2009}, case studies can be characterized based on the way they define their cases and units of analysis. This study investigates multiple MPL OSS projects, i.e., cases, and each commit and the corresponding issue fixed is a single unit of analysis.

\begin{table}[]
\caption{Programming languages examined.}
\centering
\begin{tabular}{|c|c|c|c|c|c|}
\hline
\textbf{\#} & \textbf{PL}  & \textbf{\#} & \textbf{PL} & \textbf{\#} & \textbf{PL} \\ \hline
1           & C/C++        & 7           & Haskell     & 13          & PHP         \\ \hline
2           & C\#          & 8           & Java        & 14          & Python      \\ \hline
3           & Clojure      & 9           & JavaScript  & 15          & Ruby        \\ \hline
4           & CoffeeScript & 10          & Kotlin      & 16          & Scala       \\ \hline
5           & Erlang       & 11          & Objective-C & 17          & TypeScript  \\ \hline
6           & Go           & 12          & Perl        & 18          & Swift       \\ \hline
\end{tabular}
\label{table:1}
\end{table}

\subsection{Case Selection}\label{CaseSelection}
In this study, we only investigated Apache MPL OSS projects. The reason why we used Apache projects is that the links between issues and corresponding commits tend to be well recorded in the commit messages of those projects. For selecting each case (i.e., MPL OSS project) included in our study, we applied the following inclusion criteria:

 \begin{itemize}
  \item C1: No less than 3 out of the 18 PLs listed in TABLE \ref{table:1} are used in the project. All the 18 listed PLs are general-purpose languages. Sixteen out of the 18 PLs were adopted from \cite{KoWiLo2016}, in which C and C++ are different PLs. However, we combined C and C++ into a single PL in this work since we cannot determine a header file with the extension of “.h” as a source file of C or C++ by only checking the commit records. Besides, we added two general-purpose PLs, i.e., Kotlin and Swift.
  \item C2: The source code written by the main PL is no more than 75\% of the code of the project.
  \item C3: The project has more than 2,000 commits.
  \item C4: The project has more than 20 contributors.
  \item C5: The project has more than 1,000 issues.
  \item C6: The issues of the project are tracked in JIRA.
\end{itemize}

Selection criteria C1 and C2 were set to ensure the number of MPLCs in each selected project is not too small. Criteria C3-C5 were set to ensure that the selected projects are non-trivial and the resulting dataset is big enough to be statistically analyzed. The criterion C6 was set to ensure the same format of all issues of different projects so that we can handle the issue in the same way. 

\subsection{Data Collection}\label{DataCollection}
\subsubsection{Data Items to be Collected}
\label{dataitems}
To answer the RQs, we took a commit as the unit of analysis and the data items to be collected are listed in TABLE \ref{table:2}; and to answer RQ5, we also needed to collect the data items described in TABLE \ref{table:3} of each source file, which can be extracted from the commits containing the source file. Considering all the data items to be collected except for D10 are straightforward, we only explain the definition of the entropy of the modified source files in a commit (i.e., D10) in detail \cite{Ha2009}. Suppose that the modified source files of commit $c$ are  $\{f_1,f_2,\cdots,f_n\}$, and file $f_i\left(1\leq i\leq n\right)$ was modified in $m_i$ commits during a period of time before the commit. Let $p_i = m_i/\sum_{i=1}^nm_i$. Then, the entropy $H(n) = -\sum_{i=1}^np_ilog_2 p_i$. Since the number of modified source files differs between different periods, we need to normalize the entropy to be comparable. The normalized entropy $\widetilde{H}(n) = H(n)/log_2 n$ if $n>1$, and $\widetilde{H}(n) = 0$ if $n=1$. In this study, the period is set to 60 days (including the day when commit $c$ happened), which is chosen according to \cite{LiLiLiMoLi2020}.
 
\begin{table}[]
\caption{Data items to be collected for each commit.}\label{table:2}
 \centering
\begin{tabular}{|l|l|l|l|}
\hline
\textbf{\#} & \textbf{Name} & \textbf{Description}                                                                                                               & \textbf{RQ}                                        \\ \hline
D1          & CmtID         & The hashcode of the commit.                                                                                                            & -                                                  \\ \hline
D2          & CmtDT         & \begin{tabular}[c]{@{}l@{}}The date and time when the commit \\ happened.\end{tabular}                                                 & -                                                  \\ \hline
D3          & Committer     & The committer of the commit.                                                                                                       & -                                                  \\ \hline
D4          & IsMPLC        & Whether the commit is an MPLC.                                                          & \begin{tabular}[c]{@{}l@{}}RQ1-\\ RQ5\end{tabular} \\ \hline
D5          & PL            & \begin{tabular}[c]{@{}l@{}}The programming languages used in the\\  modified source files of the commit.\\ \end{tabular}           & RQ2                                                \\ \hline
D6          & PLNo          & \begin{tabular}[c]{@{}l@{}}The number of programming languages\\  used in the modified source files of the\\  commit.\end{tabular} & RQ2                                                \\ \hline
D7          & LOCM          & \begin{tabular}[c]{@{}l@{}}The number of lines of source code\\  modified in the commit.\end{tabular}                              & RQ3                                                \\ \hline
D8          & NOFM          & \begin{tabular}[c]{@{}l@{}}The number of source files modified in \\ the commit.\end{tabular}                                      & RQ3                                                \\ \hline
D9          & NODM          & \begin{tabular}[c]{@{}l@{}}The number of directories modified in\\  the commit.\end{tabular}                                       & RQ3                                                \\ \hline
D10         & Entropy      & \begin{tabular}[c]{@{}l@{}}The normalized entropy of the modified  \\ files in the commit \cite{Ha2009}.\end{tabular}                                       & RQ3                                                \\ \hline
D11         & IssueID       & \begin{tabular}[c]{@{}l@{}}The ID of the issue fixed (i.e., mentioned) \\ in the commit if applicable.\end{tabular}                                  & \begin{tabular}[c]{@{}l@{}}RQ4,\\ RQ5\end{tabular} \\ \hline
D12         & IssueRT       & The reporting time of the issue.                                                                                                   & RQ4                                                \\ \hline
D13         & IssueCT       & \begin{tabular}[c]{@{}l@{}}The closing or resolving time of the \\ issue.\end{tabular}                                             & RQ4                                                \\ \hline
D14         & IssueType     & The type (e.g., bug) of the issue.                                                                                                 & RQ5                                                \\ \hline
\end{tabular}
\end{table}

\begin{table}[]
\caption{Data items to be collected for each source file.}\label{table:3}
\centering
\begin{tabular}{|l|l|l|l|}
\hline
\textbf{\#} & \textbf{Name} & \textbf{Description}                                                                            & \textbf{RQ} \\ \hline
D15       & Path          & The path of the source file.                                                                    & RQ5         \\ \hline
D16       & LOC           & The number of lines of code in the source file.      & RQ5         \\ \hline
D17       & NOB           & \begin{tabular}[c]{@{}l@{}}The number of bugs that the source file \\  experienced.\end{tabular} & RQ5         \\ \hline
\end{tabular}
\end{table}

\subsubsection{Data Collection Procedure}\label{datacollectionprocedure}
\begin{figure}
\centerline{\includegraphics[width=2.5in]{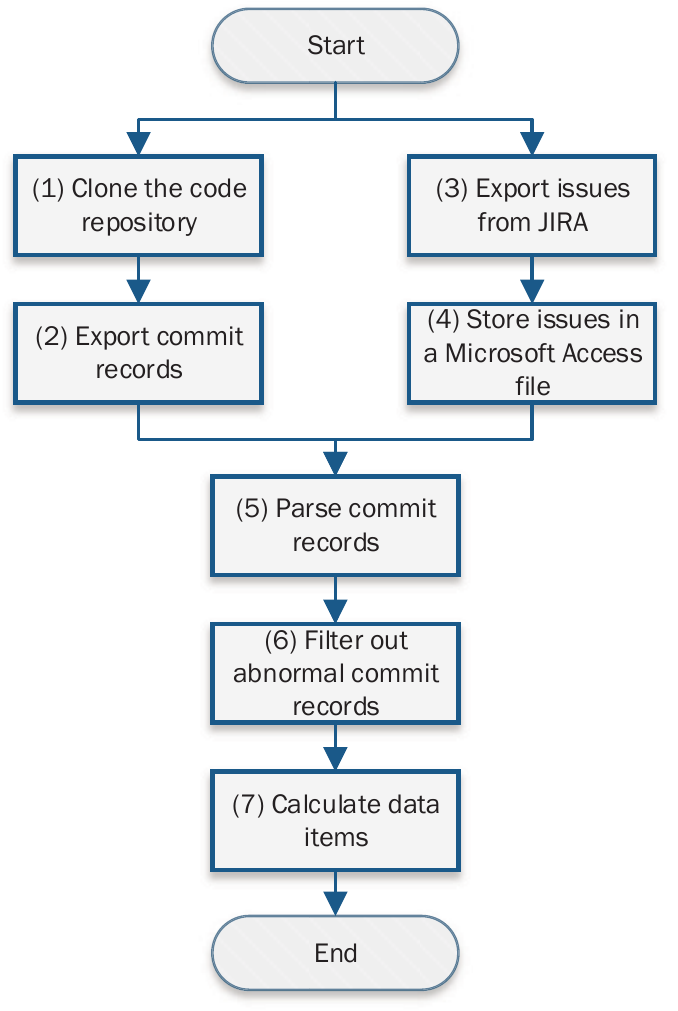}}
\caption{Procedure of data collection.}
\label{fig1}
\end{figure}

The data collection procedure for each selected project consists of seven steps (shown in Fig. \ref{fig1}). The details of the steps are described as follows.

\noindent \textbf{Step 1}: Clone the repository of the project using TortoiseGit.

\noindent \textbf{Step 2}: Extract commit records from the repository to a text file for further parsing. In this step, we only exported the commit records of the master branch and the commit records merged to the master branch, but excluded the commit records corresponding to the MERGE operations of Git. A commit record corresponding to the MERGE operation in Git is duplicate with the merged commit records, in the sense that the file changes in the MERGE commit record are the same as the file changes in the merged commit records. In addition, the committer of the MERGE commit record is different from the committers of the merged commit records. Thus, the MERGE commit record should not be included to prevent double counting code changes.

\noindent \textbf{Step 3}: Export the issue reports. We manually exported all issues of the project from JIRA – deployed by the Apache Software Foundation.

\noindent \textbf{Step 4}: Store the exported issues in Step 3 to a Microsoft Access file. 

\noindent \textbf{Step 5}: Parse commit records. If a commit is conducted to fix an issue, the committer would mention the issue ID explicitly in the commit message. The path and the number of lines of code changed of each modified source file can also be obtained in the commit record.

\noindent \textbf{Step 6}: Filter out abnormal commit records. Some commits contain changed source files with a large number of modified lines of code. For instance, if the involved source files are generated automatically, such files may be modified with tens of thousands of lines of code. Such commits should be filtered out. In this step, we filtered out commits in which more than 10,000 lines of code were modified.

\noindent \textbf{Step 7}: Calculate data items. We calculated the data items listed in TABLE \ref{table:2} and TABLE \ref{table:3}.

\begin{table*}[]
\caption{Demographic information of the selected projects.}
\centering
\begin{tabular}{|c|c|r|r|r|r|r|r|r|c|r|}
\hline
\textbf{Project} & \textbf{Name} & \multicolumn{1}{c|}{\textbf{Age (yr)}} & \multicolumn{1}{c|}{\textbf{\#LOC}} & \multicolumn{1}{c|}{\textbf{\#Commit}} & \multicolumn{1}{c|}{\textbf{\#Contributor}} & \multicolumn{1}{c|}{\textbf{\#Issue}} & \multicolumn{1}{c|}{\textbf{\#Bug}} & \multicolumn{1}{c|}{\textbf{\#PL}} & \textbf{Main PL} & \multicolumn{1}{c|}{\textbf{\%Main PL}} \\ \hline
P1               & Airavata      & 9                                      & 1,057K                              & 9,290                                  & 45                                          & 3,364                                 & 1,535                               & 5                                  & Java             & 74.7                                   \\ \hline
P2               & Ambari        & 9                                      & 1,093K                              & 24,588                                 & 134                                         & 25,261                                & 17,881                              & 11                                 & Java             & 45.9                                   \\ \hline
P3               & Arrow         & 5                                      & 635K                                & 7,575                                  & 476                                         & 10,058                                & 3,410                               & 9                                 & C/C++            & 45.0                                   \\ \hline
P4               & Avro          & 11                                     & 222K                                & 2,412                                  & 174                                         & 2,926                                 & 1,324                               & 8                                  & Java             & 45.1                                   \\ \hline
P5               & Beam          & 5                                      & 1,053K                              & 29,021                                 & 676                                         & 11,019                                & 4,608                               & 8                                  & Java             & 72.3                                   \\ \hline
P6               & Carbondata    & 4                                      & 321K                                & 4,705                                  & 171                                         & 4,019                                 & 2,199                               & 5                                  & Scala            & 57.8                                   \\ \hline
P7               & Cloudstack    & 10                                     & 912K                                & 32,644                                 & 329                                         & 10,312                                & 7,854                               & 7                                  & Java             & 59.5                                   \\ \hline
P8               & Couchdb       & 12                                     & 123K                                & 12,376                                 & 162                                         & 3,292                                 & 1,878                               & 6                                  & Erlang           & 68.0                                   \\ \hline
P9               & Dispatch      & 6                                      & 117K                                & 2,769                                  & 23                                          & 1,800                                 & 1,080                               & 5                                  & Python           & 42.3                                   \\ \hline
P10              & Ignite        & 6                                      & 2,056K                              & 27,056                                 & 241                                         & 12,727                                & 5,575                               & 8                                  & Java             & 74.6                                   \\ \hline
P11              & Impala        & 9                                      & 640K                                & 9,429                                  & 146                                         & 10,168                                & 5,630                               & 5                                 & C/C++            & 54.5                                   \\ \hline
P12              & Kafka         & 9                                      & 653K                                & 7,990                                  & 702                                         & 10,551                                & 5,809                               & 10                                 & Java             & 73.2                                   \\ \hline
P13              & Kylin         & 6                                      & 284K                                & 8,404                                  & 177                                         & 4,120                                 & 2,115                               & 6                                  & Java             & 71.3                                   \\ \hline
P14              & Ranger        & 6                                      & 327K                                & 3,371                                  & 77                                          & 3,013                                 & 2,048                               & 4                                  & Java             & 68.9                                   \\ \hline
P15              & Reef          & 8                                      & 283K                                & 3,873                                  & 68                                          & 2,063                                 & 498                                 & 6                                  & Java             & 52.9                                   \\ \hline
P16              & Spark         & 10                                     & 973K                                & 28,142                                 & 1,634                                       & 28,983                                & 12,402                              & 6                                  & Scala            & 73.6                                   \\ \hline
P17              & Subversion    & 20                                     & 860K                                & 59,809                                 & 28                                          & 4,503                                 & 3,233                               & 6                                  & C/C++            & 65.8                                   \\ \hline
P18              & Thrift        & 14                                     & 258K                                & 6,101                                  & 340                                         & 5,283                                 & 2,923                               & 14                                 & C/C++            & 33.7                                   \\ \hline
P19              & Usergrid      & 9                                      & 243K                                & 10,953                                 & 74                                          & 1,339                                 & 349                                 & 11                                 & Java             & 66.3                                   \\ \hline
P20              & Zeppelin      & 7                                      & 222K                                & 4,675                                  & 327                                         & 5,068                                 & 2,581                               & 6                                  & Java             & 59.3                                   \\ \hline
\multicolumn{2}{|c|}{\textbf{Mean}}          & \textbf{9}                                      & \textbf{617K}                                & \textbf{14,759}                                 & \textbf{300}                                         & \textbf{7,993}                                 & \textbf{4,247}                               & \textbf{7}                                  & \textbf{-}                & \textbf{60.2}                                   \\ \hline
\end{tabular}
\label{table:5}
\end{table*}

\subsection{Data Analysis}
The answers to RQ1 and RQ2 can be obtained by descriptive statistics. To answer RQ3 – RQ5, in addition to descriptive statistics, we performed Mann-Whitney U tests \cite{Fi2013} to examine if two groups are significantly different from each other. 
Since the data of the variables to be tested do not necessarily follow a specific distribution, it is reasonable to use the Mann-Whitney U test -- a non-parametric test -- in this study. The test is significant at \emph{p-value} \textless{} 0.05, which means that the tested groups have a significant difference. 

\section{Study Results}
\label{chap:study}

We collected data items described in TABLE \ref{table:2} and TABLE \ref{table:3} from 20 non-trivial Apache MPL OSS projects that were selected following the criteria presented in Section \ref{CaseSelection}. The data of the selected projects were collected around the beginning of December of 2019. TABLE \ref{table:5} shows the demographic information of the selected projects. The mean age of the projects is 9 years, the mean number of lines of code is 617K, the mean number of commits is 14,759, the mean number of contributors is 300, the mean number of issues is 7,993, and the mean number of bugs is 4,247. The number of PLs used in the projects ranges from 4 to 14, and the mean number of PLs used is 7. The percentage of code in the main PL (i.e., \%Main PL) 
of the projects ranges from 33.7\% to 74.7\%, and the mean percentage is 60.2\%. The details of the use of PLs for each project are available online\footnote{\url{https://github.com/ASSMS/ICPC2021/blob/main/PL.pdf}}. In the rest of this section, we will present the results for each research question.


\subsection{Proportion of MPLCs in the Selected Projects (RQ1)}
As shown in TABLE \ref{table:6}, the total number of commits of all projects is 205,994, the total number of MPLCs is 18,469, and thus \textbf{the percentage of MPLCs is 9.0\% when taking all projects as a whole}. Please note that the number of commits for each project in TABLE \ref{table:6} is different from the number in TABLE \ref{table:5}. This is because we only considered the commits of and the commits merged into the master branch of the project repository. 
As presented in TABLE \ref{table:6}, the proportion of MPLCs over all commits of each project ranges from 0.7\% to 41.0\%. In projects \emph{Avro} and \emph{Usegrid}, the number of MPLCs is even only around 50, much less than the other projects.

\begin{table}[]
\caption{Percentage of MPLCs in the selected projects (RQ1).}
\centering
\begin{tabular}{|c|c|r|r|r|}
\hline
\textbf{Project}   & \textbf{Name}   & \multicolumn{1}{c|}{\textbf{\#Commit}} & \multicolumn{1}{c|}{\textbf{\#MPLC}} & \multicolumn{1}{c|}{\textbf{\%MPLC}} \\ \hline
P1                 & Airavata        & 5,562                                  & 269                                  & 4.8                                  \\ \hline
P2                 & Ambari          & 19,667                                 & 1,491                                & 7.6                                  \\ \hline
P3                 & Arrow           & 4,947                                  & 1,073                                & 21.7                                 \\ \hline
P4                 & Avro            & 1,637                                  & 52                                   & 3.2                                  \\ \hline
P5                 & Beam            & 14,875                                 & 251                                  & 1.7                                  \\ \hline
P6                 & Carbondata      & 3,330                                  & 1,366                                & 41.0                                 \\ \hline
P7                 & Cloudstack      & 23,285                                 & 1,786                                & 7.7                                  \\ \hline
P8                 & Couchdb         & 7,114                                  & 560                                  & 7.9                                  \\ \hline
P9                 & Dispatch        & 2,107                                  & 381                                  & 18.1                                 \\ \hline
P10                & Ignite          & 17,565                                 & 888                                  & 5.1                                  \\ \hline
P11                & Impala          & 7,485                                  & 1,975                                & 26.4                                 \\ \hline
P12                & Kafka           & 6,915                                  & 1,029                                & 14.9                                 \\ \hline
P13                & Kylin           & 5,587                                  & 126                                  & 2.3                                  \\ \hline
P14                & Ranger          & 2,445                                  & 203                                  & 8.3                                  \\ \hline
P15                & Reef            & 2,520                                  & 163                                  & 6.5                                  \\ \hline
P16                & Spark           & 21,782                                 & 2,772                                & 12.7                                 \\ \hline
P17                & Subversion      & 44,993                                 & 3,113                                & 6.9                                  \\ \hline
P18                & Thrift          & 4,409                                  & 512                                  & 11.6                                 \\ \hline
P19                & Usergrid        & 6,791                                  & 45                                   & 0.7                                  \\ \hline
P20                & Zeppelin        & 2,978                                  & 414                                  & 13.9                                 \\ \hline
\multicolumn{2}{|c|}{\textbf{Total}} & \textbf{205,994}                       & \textbf{18,469}                      & \textbf{9.0}                         \\ \hline
\end{tabular}
\label{table:6}
\end{table}

\begin{figure}
\centerline{\includegraphics[width=3.5in]{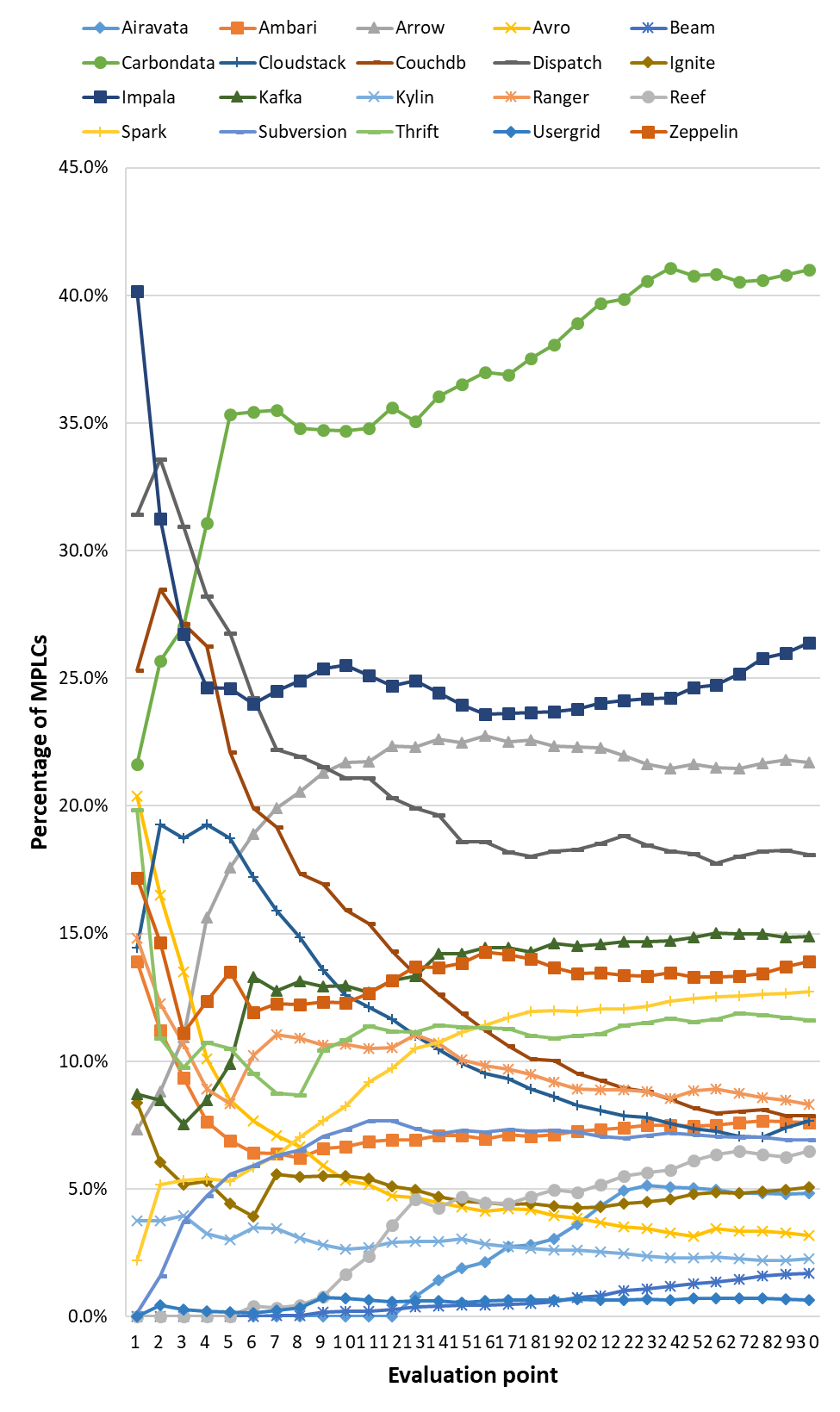}}
\caption{Trend of the proportion of MPLCs over time for each project (RQ1).}
\label{fig2}
\end{figure}

We further investigated the trend of the proportion of MPLCs in each selected project over time.  To clearly display the trends of all the 20 projects in one diagram, we selected 30 evaluation points for each project to calculate the proportion of MPLCs. At the $k$th evaluation point, we calculated the proportion of MPLCs based on the first $k$ thirtieth of all commits of the project. As shown in Fig. \ref{fig2}, \textbf{after relatively strong fluctuations in the early development stage of the projects, the proportion of MPLCs of 
most (16 out of 20, 80\%) of the projects tends to be stable in the late development stage}, while the proportion of MPLCs shows a long-term upward trend for \emph{Carbondata} and a long-term downward trend for \emph{Avro}, \emph{Cloudstack}, and \emph{Couchdb}.

\subsection{Number of PLs Used in the Source Files Modified in MPLCs (RQ2)}
\begin{figure}
\centerline{\includegraphics[width=3.7in]{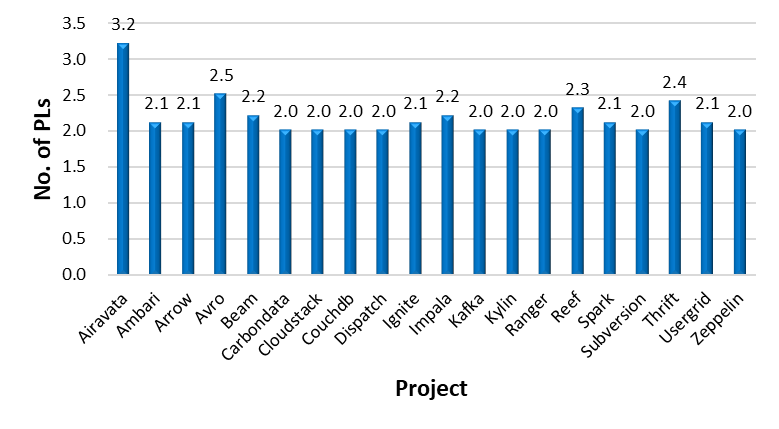}}
\caption{ Number of PLs used in the modified source files of MPLCs of the selected projects (RQ2).}
\label{fig3}
\end{figure}

Fig. \ref{fig3} shows the average number of PLs used in the source files modified in MPLCs. Among the projects, on average, project \emph{Airavata} has 3.2 PLs used in the source files that are modified in each MPLC, and this project is the only project with no less than 3.0 PLs used in the source files modified in each MPLC. Most (16 out of 20, 80\%) of the projects have around 2.0 (i.e., 2.0-2.2) PLs for each MPLC on average. 

We further explored how the number of PLs used distributes over MPLCs, and the results are shown in TABLE \ref{table:7}. In this table, \emph{\#Ci} denotes the number of commits in which the modified source files are written in \emph{i} PLs, \emph{\%Ci} denotes the percentage of \emph{\#Ci} over \emph{\#MPLC}, and \emph{\#C5+} denotes the number of commits in which the modified source files are written in 5 or more PLs. As shown in TABLE \ref{table:7}, taking all the projects as a whole, 91.7\%, 7.1\%, and 1.0\% of the MPLCs involve source files written in 2, 3, and 4 PLs, respectively; and only 0.2\% of the MPLCs involve source files written in 5 or more PLs. This means that \textbf{most MPLCs involve source files written in only 2 PLs, and it is not common for MPLCs to modify source files in more than 4 PLs.}

TABLE \ref{table:7} also shows that most (14 out of 20, 70\%) of the projects do not have MPLCs with 5 or more PLs, and about one third (7 out of 20) of the projects do not have MPLCs with more than 3 PLs.

\begin{table*}[]
\caption{Distribution of the number of PLs used in the modified source files of MPLCs of the selected projects (RQ2).}
\centering
\setlength{\tabcolsep}{3mm}{
\begin{tabular}{|c|c|r|r|r|r|r|r|r|r|r|}
\hline
\textbf{Project} & \textbf{Name} & \multicolumn{1}{c|}{\textbf{\#MPLC}} & \multicolumn{1}{c|}{\textbf{\#C2}} & \multicolumn{1}{c|}{\textbf{\%C2}} & \multicolumn{1}{c|}{\textbf{\#C3}} & \multicolumn{1}{c|}{\textbf{\%C3}} & \multicolumn{1}{c|}{\textbf{\#C4}} & \multicolumn{1}{c|}{\textbf{\%C4}} & \multicolumn{1}{c|}{\textbf{\#C5+}} & \multicolumn{1}{c|}{\textbf{\%C5+}} \\ \hline
P1               & Airavata      & 269                                  & 59                                 & 21.9                               & 86                                 & 32.0                               & 124                                & 46.1                               & 0                                   & 0.0                                 \\ \hline
P2               & Ambari        & 1,491                                & 1,383                              & 92.8                               & 105                                & 7.0                                & 2                                  & 0.1                                & 1                                   & 0.1                                 \\ \hline
P3               & Arrow         & 1,073                                & 1,013                              & 94.4                               & 59                                 & 5.5                                & 1                                  & 0.1                                & 0                                   & 0.0                                 \\ \hline
P4               & Avro          & 52                                   & 39                                 & 75.0                               & 9                                  & 17.3                               & 1                                  & 1.9                                & 3                                   & 5.8                                 \\ \hline
P5               & Beam          & 251                                  & 212                                & 84.5                               & 39                                 & 15.5                               & 0                                  & 0.0                                & 0                                   & 0.0                                 \\ \hline
P6               & Carbondata    & 1,366                                & 1,365                              & 99.9                               & 1                                  & 0.1                                & 0                                  & 0.0                                & 0                                   & 0.0                                 \\ \hline
P7               & Cloudstack    & 1,786                                & 1,706                              & 95.5                               & 79                                 & 4.4                                & 1                                  & 0.1                                & 0                                   & 0.0                                 \\ \hline
P8               & Couchdb       & 560                                  & 547                                & 97.7                               & 10                                 & 1.8                                & 3                                  & 0.5                                & 0                                   & 0.0                                 \\ \hline
P9               & Dispatch      & 381                                  & 378                                & 99.2                               & 3                                  & 0.8                                & 0                                  & 0.0                                & 0                                   & 0.0                                 \\ \hline
P10              & Ignite        & 888                                  & 788                                & 88.7                               & 94                                 & 10.6                               & 5                                  & 0.6                                & 1                                   & 0.1                                 \\ \hline
P11              & Impala        & 1,975                                & 1,606                              & 81.3                               & 368                                & 18.6                               & 1                                  & 0.1                                & 0                                   & 0.0                                 \\ \hline
P12              & Kafka         & 1,029                                & 992                                & 96.4                               & 37                                 & 3.6                                & 0                                  & 0.0                                & 0                                   & 0.0                                 \\ \hline
P13              & Kylin         & 126                                  & 125                                & 99.2                               & 1                                  & 0.8                                & 0                                  & 0.0                                & 0                                   & 0.0                                 \\ \hline
P14              & Ranger        & 203                                  & 199                                & 98.0                               & 4                                  & 2.0                                & 0                                  & 0.0                                & 0                                   & 0.0                                 \\ \hline
P15              & Reef          & 163                                  & 115                                & 70.6                               & 47                                 & 28.8                               & 1                                  & 0.6                                & 0                                   & 0.0                                 \\ \hline
P16              & Spark         & 2,772                                & 2,491                              & 89.9                               & 277                                & 10.0                               & 4                                  & 0.1                                & 0                                   & 0.0                                 \\ \hline
P17              & Subversion    & 3,113                                & 3,039                              & 97.6                               & 53                                 & 1.7                                & 17                                 & 0.5                                & 4                                   & 0.1                                 \\ \hline
P18              & Thrift        & 512                                  & 436                                & 85.2                               & 31                                 & 6.1                                & 14                                 & 2.7                                & 31                                  & 6.1                                 \\ \hline
P19              & Usergrid      & 45                                   & 43                                 & 95.6                               & 1                                  & 2.2                                & 0                                  & 0.0                                & 1                                   & 2.2                                 \\ \hline
P20              & Zeppelin      & 414                                  & 396                                & 95.7                               & 16                                 & 3.9                                & 2                                  & 0.5                                & 0                                   & 0.0                                 \\ \hline
\multicolumn{2}{|c|}{\textbf{Total}}    & \textbf{18,469}                      & \textbf{16,932}                    & \textbf{91.7}                      & \textbf{1,320}                     & \textbf{7.1}                       & \textbf{176}                       & \textbf{1.0}                       & \textbf{41}                         & \textbf{0.2}                        \\ \hline
\end{tabular}
}

\label{table:7}
\end{table*}

\subsection{Change Complexity of MPLCs (RQ3)}
Change complexity can be measured by the number of lines of code modified (LOCM), number of source files modified (NOFM), number of directories modified (NODM), and entropy of source files modified (Entropy) \cite{LiLiLiMoLi2020}. In TABLE \ref{table:9}, \emph{AveM} and \emph{AveN} denote the average value of the corresponding change complexity measure of MPLCs and non-MPLCs, respectively; \emph{\%Diff} denotes the percentage of the difference between \emph{AveM} and \emph{AveN} (i.e., $\%Diff=((\emph{AveM}-\emph{AveN})/\emph{AveN})\times{100\%}$)). As shown in TABLE \ref{table:9}, for all the projects, on average, these four change complexity measures of MPLCs are much larger than those of non-MPLCs, respectively. Specifically, on average, the LOCM, NOFM, and NOFM of MPLCs are larger than those of non-MPLCs by more than 100.0\% for most (85\%+) of the projects, and the Entropy of MPLCs is larger than that of non-MPLCs by 40.6\% at least. We further ran Mann-Whitney U tests on the four measures of MPLCs and non-MPLCs for each project, and the \emph{p-value} for each measure of each project is less than 0.001 (except for the \emph{p-value} for the Entropy of project P4, which is 0.015) as shown in TABLE \ref{table:9}. This indicates that all the four measures of MPLCs of each project are significantly larger than the measures for non-MPLCs, respectively. In other words, \textbf{the change complexity of MPLCs is significantly higher than that of non-MPLCs for each selected project.}

\begin{table*}[]
\caption{Change complexity of MPLCs and non-MPLCs (RQ3).}
\centering
\setlength{\tabcolsep}{1mm}{\begin{tabular}{|c|c|r|r|r|r|r|r|r|r|r|r|r|r|r|r|r|r|}
\hline
\multirow{2}{*}{\textbf{Project}} & \multirow{2}{*}{\textbf{Name}} & \multicolumn{4}{c|}{\textbf{LOCM}}                                                                                                                    & \multicolumn{4}{c|}{\textbf{NOFM}}                                                                                                                    & \multicolumn{4}{c|}{\textbf{NODM}}                                                                                                                    & \multicolumn{4}{c|}{\textbf{Entropy}}                                                                                                                 \\ \cline{3-18} 
                                  &                                & \multicolumn{1}{c|}{\textbf{AveM}} & \multicolumn{1}{c|}{\textbf{AveN}} & \multicolumn{1}{c|}{\textbf{\%Diff}} & \multicolumn{1}{c|}{\textit{\textbf{p-value}}} & \multicolumn{1}{c|}{\textbf{AveM}} & \multicolumn{1}{c|}{\textbf{AveN}} & \multicolumn{1}{c|}{\textbf{\%Diff}} & \multicolumn{1}{c|}{\textit{\textbf{p-value}}} & \multicolumn{1}{c|}{\textbf{AveM}} & \multicolumn{1}{c|}{\textbf{AveN}} & \multicolumn{1}{c|}{\textbf{\%Diff}} & \multicolumn{1}{c|}{\textit{\textbf{p-value}}} & \multicolumn{1}{c|}{\textbf{AveM}} & \multicolumn{1}{c|}{\textbf{AveN}} & \multicolumn{1}{c|}{\textbf{\%Diff}} & \multicolumn{1}{c|}{\textit{\textbf{p-value}}} \\ \hline
P1                                & Airavata                       & 1,645                           & 313                             & 425.6                            & \textless{}0.001                               & 31                              & 5                               & 520.0                            & \textless{}0.001                               & 12                              & 3                               & 300.0                            & \textless{}0.001                               & 0.93                            & 0.50                            & 86.0                             & \textless{}0.001                               \\ \hline
P2                                & Ambari                         & 541                             & 187                             & 189.3                            & \textless{}0.001                               & 13                              & 5                               & 160.0                            & \textless{}0.001                               & 7                               & 3                               & 133.3                            & \textless{}0.001                               & 0.88                            & 0.58                            & 51.7                             & \textless{}0.001                               \\ \hline
P3                                & Arrow                          & 517                             & 262                             & 97.3                             & \textless{}0.001                               & 10                              & 6                               & 66.7                             & \textless{}0.001                               & 4                               & 2                               & 100.0                            & \textless{}0.001                               & 0.91                            & 0.64                            & 42.2                             & \textless{}0.001                               \\ \hline
P4                                & Avro                           & 780                             & 264                             & 195.5                            & \textless{}0.001                               & 32                              & 6                               & 433.3                            & \textless{}0.001                               & 7                               & 3                               & 133.3                            & \textless{}0.001                               & 0.97                            & 0.69                            & 40.6                             & 0.015                                          \\ \hline
P5                                & Beam                           & 578                             & 226                             & 155.8                            & \textless{}0.001                               & 20                              & 6                               & 233.3                            & \textless{}0.001                               & 8                               & 3                               & 166.7                            & \textless{}0.001                               & 0.90                            & 0.57                            & 57.9                             & \textless{}0.001                               \\ \hline
P6                                & Carbondata                     & 636                             & 264                             & 140.9                            & \textless{}0.001                               & 15                              & 5                               & 200.0                            & \textless{}0.001                               & 10                              & 3                               & 233.3                            & \textless{}0.001                               & 0.89                            & 0.59                            & 50.8                             & \textless{}0.001                               \\ \hline
P7                                & Cloudstack                     & 411                             & 156                             & 163.5                            & \textless{}0.001                               & 7                               & 4                               & 75.0                             & \textless{}0.001                               & 5                               & 2                               & 150.0                            & \textless{}0.001                               & 0.87                            & 0.34                            & 155.9                            & \textless{}0.001                               \\ \hline
P8                                & Couchdb                        & 266                             & 95                              & 180.0                            & \textless{}0.001                               & 5                               & 2                               & 150.0                            & \textless{}0.001                               & 2                               & 1                               & 100.0                            & \textless{}0.001                               & 0.87                            & 0.27                            & 222.2                            & \textless{}0.001                               \\ \hline
P9                                & Dispatch                       & 456                             & 145                             & 214.5                            & \textless{}0.001                               & 9                               & 3                               & 200.0                            & \textless{}0.001                               & 4                               & 2                               & 100.0                            & \textless{}0.001                               & 0.90                            & 0.42                            & 114.3                            & \textless{}0.001                               \\ \hline
P10                               & Ignite                         & 1,019                           & 292                             & 249.0                            & \textless{}0.001                               & 40                              & 7                               & 471.4                            & \textless{}0.001                               & 16                              & 4                               & 300.0                            & \textless{}0.001                               & 0.90                            & 0.53                            & 69.8                             & \textless{}0.001                               \\ \hline
P11                               & Impala                         & 455                             & 156                             & 191.7                            & \textless{}0.001                               & 12                              & 4                               & 200.0                            & \textless{}0.001                               & 5                               & 2                               & 150.0                            & \textless{}0.001                               & 0.90                            & 0.53                            & 69.8                             & \textless{}0.001                               \\ \hline
P12                               & Kafka                          & 709                             & 202                             & 251.0                            & \textless{}0.001                               & 18                              & 5                               & 260.0                            & \textless{}0.001                               & 9                               & 3                               & 200.0                            & \textless{}0.001                               & 0.91                            & 0.57                            & 59.6                             & \textless{}0.001                               \\ \hline
P13                               & Kylin                          & 274                             & 193                             & 42.0                             & \textless{}0.001                               & 7                               & 5                               & 40.0                             & \textless{}0.001                               & 5                               & 3                               & 66.7                             & \textless{}0.001                               & 0.86                            & 0.51                            & 68.6                             & \textless{}0.001                               \\ \hline
P14                               & Ranger                         & 563                             & 262                             & 114.9                            & \textless{}0.001                               & 11                              & 5                               & 120.0                            & \textless{}0.001                               & 7                               & 3                               & 133.3                            & \textless{}0.001                               & 0.92                            & 0.51                            & 80.4                             & \textless{}0.001                               \\ \hline
P15                               & Reef                           & 530                             & 254                             & 108.7                            & \textless{}0.001                               & 23                              & 8                               & 187.5                            & \textless{}0.001                               & 9                               & 4                               & 125.0                            & \textless{}0.001                               & 0.92                            & 0.61                            & 50.8                             & \textless{}0.001                               \\ \hline
P16                               & Spark                          & 404                             & 126                             & 220.6                            & \textless{}0.001                               & 11                              & 4                               & 175.0                            & \textless{}0.001                               & 6                               & 3                               & 100.0                            & \textless{}0.001                               & 0.88                            & 0.54                            & 63.0                             & \textless{}0.001                               \\ \hline
P17                               & Subversion                     & 235                             & 75                              & 213.3                            & \textless{}0.001                               & 7                               & 2                               & 250.0                            & \textless{}0.001                               & 3                               & 1                               & 200.0                            & \textless{}0.001                               & 0.86                            & 0.25                            & 244.0                            & \textless{}0.001                               \\ \hline
P18                               & Thrift                         & 451                             & 128                             & 252.3                            & \textless{}0.001                               & 8                               & 3                               & 166.7                            & \textless{}0.001                               & 4                               & 1                               & 300.0                            & \textless{}0.001                               & 0.90                            & 0.38                            & 136.8                            & \textless{}0.001                               \\ \hline
P19                               & Usergrid                       & 555                             & 259                             & 114.3                            & \textless{}0.001                               & 10                              & 5                               & 100.0                            & \textless{}0.001                               & 4                               & 3                               & 33.3                             & \textless{}0.001                               & 0.92                            & 0.50                            & 84.0                             & \textless{}0.001                               \\ \hline
P20                               & Zeppelin                       & 541                             & 167                             & 224.0                            & \textless{}0.001                               & 10                              & 4                               & 150.0                            & \textless{}0.001                               & 6                               & 2                               & 200.0                            & \textless{}0.001                               & 0.90                            & 0.46                            & 95.7                             & \textless{}0.001                               \\ \hline
\end{tabular}}
\label{table:9}
\end{table*}

\subsection{Open Time of Issues Fixed in MPLCs (RQ4)}

We studied the open time (i.e., the time from when an issue report is created to when the issue is resolved) of issues fixed in MPLCs, and the results are shown in TABLE \ref{table:8}, in which \emph{AveOTM} and \emph{AveOTN} denote the average open time of issues fixed in MPLCs and non-MPLCs respectively, \emph{\%Diff} denotes the difference between \emph{AveOTM} and \emph{AveOTN} (i.e., $\%Diff=((AveOTM-AveOTN)/AveOTN)\times{100\%}$), and \emph{p-value} denotes the result of the Mann-Whitney U test on the open time of issues fixed in MPLCs and non-MPLCs. As shown in TABLE \ref{table:8}, 16 out of 20 (80\%) projects have longer open time of issues fixed in MPLCs than that of issues fixed in non-MPLCs, and the other 4 projects (which corresponding cells of column \emph{p-value} are filled in grey) do not show a significant difference between the open time of issues fixed in MPLCs and that of issues fixed in non-MPLCs. The average open time of issues fixed in MPLCs is 8.0\% to 124.7\% longer than that of issues fixed in non-MPLCs for the 16 projects. It indicates that \textbf{issues fixed in MPLCs likely take longer to be resolved than issues fixed in non-MPLCs.}

\begin{table}[]
\caption{Average open time of issues fixed in MPLCs of the selected projects (RQ4).}
\centering
\setlength{\tabcolsep}{1.5mm}{
\begin{tabular}{|c|c|r|r|r|r|}
\hline
\textbf{Proj.} & \textbf{Name} & \multicolumn{1}{c|}{\textbf{AveOTM(day)}} & \multicolumn{1}{c|}{\textbf{AveOTN(day)}} & \multicolumn{1}{c|}{\textbf{\%Diff}} & \multicolumn{1}{c|}{\textit{\textbf{p-value}}} \\ \hline
P1               & Airavata      & 95.22                                  & 58.95                                  & 61.5                              & 0.005                                          \\ \hline
P2               & Ambari        & 13.52                                  & 8.13                                   & 66.3                              & \textless{}0.001                               \\ \hline
P3               & Arrow         & 55.83                                  & 38.94                                  & 43.4                              & 0.000                                          \\ \hline
P4               & Avro          & 60.00                                  & 128.16                                 & -53.2                             & \cellcolor[HTML]{C0C0C0}0.225                  \\ \hline
P5               & Beam          & 96.08                                  & 87.59                                  & 9.7                               & 0.007                                          \\ \hline
P6               & Carbondata    & 24.07                                  & 22.28                                  & 8.0                               & \textless{}0.001                               \\ \hline
P7               & Cloudstack    & 113.71                                 & 50.60                                  & 124.7                             & \textless{}0.001                               \\ \hline
P8               & Couchdb       & 165.91                                 & 154.39                                 & 7.5                               & \cellcolor[HTML]{C0C0C0}0.792                  \\ \hline
P9               & Dispatch      & 55.82                                  & 34.30                                  & 62.7                              & \textless{}0.001                               \\ \hline
P10              & Ignite        & 71.11                                  & 56.03                                  & 26.9                              & \textless{}0.001                               \\ \hline
P11              & Impala        & 160.51                                 & 78.66                                  & 104.1                             & \textless{}0.001                               \\ \hline
P12              & Kafka         & 98.26                                  & 67.84                                  & 44.8                              & \textless{}0.001                               \\ \hline
P13              & Kylin         & 123.40                                 & 61.64                                  & 100.2                             & 0.006                                          \\ \hline
P14              & Ranger        & 37.05                                  & 21.44                                  & 72.8                              & \textless{}0.001                               \\ \hline
P15              & Reef          & 41.99                                  & 29.24                                  & 43.6                              & \textless{}0.001                               \\ \hline
P16              & Spark         & 59.55                                  & 40.73                                  & 46.2                              & \textless{}0.001                               \\ \hline
P17              & Subversion    & 711.54                                 & 704.09                                 & 1.1                               & \cellcolor[HTML]{C0C0C0}0.388                  \\ \hline
P18              & Thrift        & 159.95                                 & 119.91                                 & 33.4                              & \textless{}0.001                               \\ \hline
P19              & Usergrid      & 69.42                                  & 83.38                                  & -16.7                             & \cellcolor[HTML]{C0C0C0}0.811                  \\ \hline
P20              & Zeppelin      & 48.30                                  & 35.64                                  & 35.5                              & \textless{}0.001                               \\ \hline
\end{tabular}
}

\label{table:8}
\end{table}

\subsection{Bug proneness of source files modified in MPLCs (RQ5)}
Bug proneness of a source file can be measured by defect density (DD) of the file, i.e., NOB/LOC of the file \cite{LiLiLi2017}. We calculated the defect density of the source files modified in MPLCs and that of the source files modified only in non-MPLCs. The results are shown in TABLE \ref{table:10}, in which \emph{\#FileAll} denotes the number of all source files of a project, \emph{Ave. LOC} denotes the average number of lines of code of all source files of a project, \emph{\#FileM} denotes the number of source files modified in MPLCs, \emph{\#FileN} denotes the number of source files modified only in non-MPLCs, \emph{DDM} denotes the defect density of a source file modified in MPLCs, \emph{DDN} denotes the defect density of a source file modified only in non-MPLCs, and \emph{\%Diff} denotes the percentage of the difference between \emph{Ave. DDM} and \emph{Ave. DDN} (i.e., $\%Diff=((\emph{Ave. DDM}-\emph{Ave. DDN})/\emph{Ave. DDN})\times{100\%}$)). We ran the Mann-Whitney U test to compare the \emph{DDM} and \emph{DDN} of source files in each project, and the \emph{p-value} is shown in TABLE \ref{table:10}. As we can see from the table, \emph{DDM} is significantly larger than \emph{DDN} with \emph{p-value} \textless{} 0.05 in 16 out of 20 (80\%) projects, and the difference between \emph{DDM} and \emph{DDN} ranges from 17.2\% to 3681.8\%; \emph{DDM} is significantly smaller than \emph{DDN} with \emph{p-value} \textless{} 0.05 in 2 projects (marked with * following the corresponding \emph{p-values} in TABLE \ref{table:10}); and there is no significant difference between \emph{DDM} and \emph{DDN} in the other 2 projects (with \emph{p-value} \textgreater{} 0.05, the corresponding cells of column \emph{p-value} are filled in grey in TABLE \ref{table:10}). In other words, \textbf{the defect density of source files that have been modified in MPLCs is likely larger than the defect density of source files that have never been modified in MPLCs.}

\begin{table*}[]
\caption{Defect density of source files modified in MPLCs and non-MPLCs (RQ5).}
\centering
\begin{tabular}{|c|c|r|r|r|r|r|r|r|r|}
\hline
\textbf{Project} & \textbf{Name} & \multicolumn{1}{c|}{\textbf{\#FileAll}} & \multicolumn{1}{c|}{\textbf{Ave. LOC}} & \multicolumn{1}{c|}{\textbf{\#FileM}} & \multicolumn{1}{c|}{\textbf{\#FileN}} & \multicolumn{1}{c|}{\textbf{Ave. DDM}} & \multicolumn{1}{c|}{\textbf{Ave. DDN}} & \multicolumn{1}{c|}{\textbf{\%Diff}} & \multicolumn{1}{c|}{\textit{\textbf{p-value}}} \\ \hline
P1               & Airavata      & 993                                       & 1,065                             & 450                                     & 543                                     & 0.00174                             & 0.00325                             & -46.5                             & \cellcolor[HTML]{C0C0C0}0.381                  \\ \hline
P2               & Ambari        & 5,645                                     & 194                               & 2,272                                   & 3,373                                   & 0.03636                             & 0.01848                             & 96.8                              & \textless{}0.001                               \\ \hline
P3               & Arrow         & 2,851                                     & 223                               & 1,757                                   & 1,094                                   & 0.01041                             & 0.00604                             & 72.4                              & \textless{}0.001                               \\ \hline
P4               & Avro          & 1,134                                     & 196                               & 976                                     & 158                                     & 0.00775                             & 0.00340                             & 127.9                             & \textless{}0.001                               \\ \hline
P5               & Beam          & 5,812                                     & 181                               & 2,344                                   & 3,468                                   & 0.01034                             & 0.00325                             & 218.2                             & \textless{}0.001                               \\ \hline
P6               & Carbondata    & 1,484                                     & 216                               & 1,244                                   & 240                                     & 0.00770                             & 0.00290                             & 165.5                             & \textless{}0.001                               \\ \hline
P7               & Cloudstack    & 2,395                                     & 381                               & 1,346                                   & 1,049                                   & 0.00256                             & 0.00247                             & 3.6                               & \cellcolor[HTML]{C0C0C0}0.071                  \\ \hline
P8               & Couchdb       & 557                                       & 221                               & 173                                     & 384                                     & 0.00038                             & 0.00009                             & 322.2                             & 0.007                                          \\ \hline
P9               & Dispatch      & 332                                       & 351                               & 261                                     & 71                                      & 0.02112                             & 0.00517                             & 308.5                             & \textless{}0.001                               \\ \hline
P10              & Ignite        & 12,270                                    & 168                               & 4,924                                   & 7,346                                   & 0.01085                             & 0.01212                             & -10.5                             & \textless{}0.001*                               \\ \hline
P11              & Impala        & 2,364                                     & 271                               & 1,784                                   & 580                                     & 0.01549                             & 0.00486                             & 218.7                             & \textless{}0.001                               \\ \hline
P12              & Kafka         & 3,341                                     & 195                               & 2,365                                   & 976                                     & 0.01395                             & 0.01190                             & 17.2                              & \textless{}0.001                               \\ \hline
P13              & Kylin         & 1,831                                     & 155                               & 328                                     & 1,503                                   & 0.01165                             & 0.00784                             & 48.6                              & \textless{}0.001                               \\ \hline
P14              & Ranger        & 1,480                                     & 221                               & 579                                     & 901                                     & 0.01861                             & 0.01394                             & 33.5                              & \textless{}0.001                               \\ \hline
P15              & Reef          & 3,590                                     & 79                                & 499                                     & 3,091                                   & 0.01271                             & 0.00421                             & 201.9                             & \textless{}0.001                               \\ \hline
P16              & Spark         & 4,854                                     & 201                               & 3,759                                   & 1,095                                   & 0.01499                             & 0.00711                             & 110.8                             & \textless{}0.001                               \\ \hline
P17              & Subversion    & 1,571                                     & 548                               & 1,296                                   & 275                                     & 0.00416                             & 0.00011                             & 3,681.8                            & \textless{}0.001                               \\ \hline
P18              & Thrift        & 1,322                                     & 195                               & 839                                     & 483                                     & 0.01307                             & 0.00742                             & 76.1                              & \textless{}0.001                               \\ \hline
P19              & Usergrid      & 1,884                                     & 129                               & 128                                     & 1,756                                   & 0.00022                             & 0.00031                             & -29.0                             & 0.028*                                          \\ \hline
P20              & Zeppelin      & 1,197                                     & 185                               & 716                                     & 481                                     & 0.00945                             & 0.00445                             & 112.4                             & \textless{}0.001                               \\ \hline
\end{tabular}
\label{table:10}
\end{table*}

\section{Discussion}
\label{chap:discus}

In this section, we interpret the results of the study according to the RQs and discuss the implications of the results for both practitioners and researchers. 
\subsection{Interpretation of Study Results}
\emph{\textbf{RQ1}}: Only 9.0\% of the commits are MPLCs when taking all projects as a whole, and the other 91.0\% of the commits are non-MPLCs. It indicates that developers tend to make mono-language changes despite the MPL development. However, from the perspective of individual projects, the proportion of MPLCs may differ greatly. As we can see from TABLE  \ref{table:5} and TABLE \ref{table:6}, a greater number of PLs used and a lower percentage of code in the main PL do not necessarily mean a larger proportion of MPLCs in a project. One possible reason is that design quality may play an important role. For instance, higher modularity may reduce the likelihood of MPLCs.

Although the proportion of MPLCs of the selected projects differs from one to one, the proportion of MPLCs goes to a relatively stable level for most (80\%) of the selected projects. This is an interesting phenomenon, which may indicate certain balanced status in the development, e.g., stable architecture design. It is worth further investigation to explore what factors play dominant roles in this phenomenon.

\emph{\textbf{RQ2}}: Most of MPLCs involve source files written in two PLs, which may be a natural choice of developers. Source files in more PLs to be modified in a commit may lead to higher complexity of the code change, which requires more comprehensive consideration of the potential influence of the code change on the quality of the software systems. However, there lack effective tools to automatically analyze change impact in an MPL context, and MPL code analysis still remains a rather challenging problem \cite{ShMiAb2019}.

\emph{\textbf{RQ3}}: Change complexity of MPLCs is significantly higher than change complexity of non-MPLCs, which is not surprising. MPLCs involve source files written in multiple PLs, and source files in different PLs are usually distributed over different components. Therefore, MPLCs tend to exert relatively global impact on the software system, and the change complexity of MPLCs is likely to be higher. In addition, the results of all the four change complexity measures (i.e., LOCM, NOFM, NODM, and Entropy) are perfectly consistent, which increases the confidence in the finding that changes in MPLCs are more complex than changes in non-MPLCs. Finally, the information on whether a code change is an MPLC can facilitate effort estimation in project management. 

\emph{\textbf{RQ4}}: The results on the open time of issues show that issues fixed in MPLCs likely take longer to be resolved than issues fixed in non-MPLCs. The open time of issues generally depends on two factors: the priority of issues and the difficulty of resolving issues. We looked into whether there is a significant difference on the priority of issues fixed in MPLCs and non-MPLCs, and found that there is no significant difference between the priority of issues fixed in MPLCs and non-MPLCs for most (12/20) of the selected projects. The results on the issue priority are not presented in this paper due to the space limit, but have been available online\footnote{\url{https://github.com/ASSMS/ICPC2021/blob/main/issuepriority.pdf}}. Thus, the main reason for issues fixed in MPLCs taking longer to be resolved may be that such issues are more difficult to be fixed, which is evidenced by the results on RQ3. 

\emph{\textbf{RQ5}}: Source files modified in MPLCs are likely to be more bug-prone, thus, the proportion of source files that have been modified in MPLCs can be used as an indicator of risk of bug introduction in MPL software systems. Source files in various PLs modified in an MPLC indicate that these source files are linked together due to (in)direct dependencies. In addition, source files in different PLs are communicated through dedicated mechanisms (e.g., JNI) and there lack cross-language analysis tools, which increases the difficulty of bug fixing and consequently results in higher bug proneness of the involved source files in MPLCs.

\subsection{Implications for Practitioners}
\textbf{Prevent too many MPLCs by architecture design of MPL software systems.} Since the change complexity of MPLCs is significantly higher than non-MPLCs, too many MPLCs happening in an MPL system will result in a considerable increase of effort for making changes to the system. An MPLC tends to be at the architecture level in light of that multiple components are modified in the MPLC. Thus, it is wise to design a more maintainable architecture for an MPL system. 
When a relatively high proportion of commits are MPLCs in an MPL system, there is a necessity to assess the maintainability (especially modularity) of the architecture of this system \cite{MoCaKaXiFe2016,WoCaKiDa2011}, and then to improve the architecture design through e.g., refactoring.

\textbf{Pay special attention to source files modified in MPLCs.} As the results of RQ5 revealed, the source files modified in MPLCs are likely to have a higher defect density than that of source files only modified in non-MPLCs. Therefore, practitioners should pay more attention to the former. For instance, designers may improve the modularity of source files modified in MPLCs, in order to lower the likelihood of such source files being modified together; developers need to investigate deeper on the impact of MPLCs; and developers and testers can invest more effort to test such source files.

\subsection{Implications for Researchers}
\textbf{Take MPLCs into account when constructing defect prediction models for MPL software systems.} Since whether the source files have been modified in MPLCs plays a role in the defect density of source files, it is reasonable to take MPLCs as a factor in defect prediction for MPL systems.

\textbf{Investigate the factors that influence the proportion of MPLCs in MPL software systems.} The change complexity of MPLCs is much higher than that of non-MPLCs, which implies that MPLCs will greatly increase the development cost of MPL software systems. Therefore, it is necessary to keep the proportion of MPLCs of an MPL system under a reasonable level. However, it still remains unclear what factors contribute to a relatively high proportion of MPLCs in an MPL system, which is an interesting research question.


\textbf{Need further studies on the relationship between MPLCs and software architecture.} Although intuitively MPLCs are more related to the changes at the architecture level, there still lacks evidence on how MPLCs relate to software architecture and verse vice. We believe that this could be a promising research topic to be further explored. 

\section{Threats to Validity}
\label{chap:threats}

There are several threats to the validity of the study results. We discuss these threats according to the guidelines in \cite{RuHo2009}. Please note that internal validity is not discussed, since we do not study causal relationships.

\subsection{Construct Validity}
Construct validity is concerned with whether the values of the variables (listed in TABLE \ref{table:2} and TABLE \ref{table:3}) we obtained are in line with the real values that we expected. A potential threat to construct validity is that not all issues resolved are linked to corresponding commits. Due to different developer habits and development cultures, committers may not explicitly mention the ID of the issue resolved in corresponding commit message, which may negatively affect the representativeness of the collected issues and further influence the accuracy of defect density and the time taken to resolve issues. Through our analysis (the analysis results are not shown in this paper due to its deviation from the focus of this paper), we confirmed that the committers who explicitly mention the issue ID do not come from a small group of specific developers. Therefore, this threat is to some extent mitigated. 


\subsection{External Validity}
External validity is concerned with the generalizability of the study results. First, a potential threat to external validity is whether the selected projects are representative enough. As presented in Section \ref{CaseSelection}, we applied a set of criteria to select projects. We tried to include as many as possible the Apache projects that meet the selection criteria. Furthermore, the selected projects cover different application domains, and differ in code repository size and development duration. This indicates improved representativeness of the selected projects.

Second, another threat is that only Apache MPL OSS projects were selected. The number of available projects is relatively small, which may reduce the generalizability of the study results. 


Finally, since only 18 PLs are considered in this study, the findings and the conclusions drawn are only valid for projects using these PLs. Since only OSS projects were selected, we cannot generalize the findings and conclusions to closed source software projects.

\subsection{Reliability}
Reliability refers to whether the study yields the same results when it is replicated by other researchers. A potential threat is related to the implementation of related software tools for data collection. The tools were mainly implemented by the third author, and the code of the key functionalities had been regularly reviewed by the first and second authors. In addition, sufficient tests were performed to ensure the correctness of the calculation of data items. Hence, the threat to reliability had been alleviated.


\section{Conclusions and Future Work}
\label{conclusions}

The phenomenon of MPLCs is prevalent in modern software system development. To our knowledge, this phenomenon has not been explored yet. In light of the potential influence of MPLCs on development difficulty and software quality, we conducted an empirical study to understand the state of MPLCs, their change complexity, as well as their impact on open time of issues and bug proneness of source files in real-life software projects.  

Following a set of predefined criteria, we selected 20 non-trivial Apache MPL OSS projects, in which 205,994 commits (including 18,469 MPLCs) were analyzed. The main findings are that:

 \begin{itemize}
  \item The proportion of MPLCs for the selected projects ranges from 0.7\% to 41.0\%, and 9.0\% of the commits are MPLCs when taking all projects as a whole. The proportion of MPLCs goes to a relatively stable level for 80\% of the selected projects.
  \item In most of the selected projects, the average number of PLs used in each MPLC is around 2.0. Particularly, when taking all selected projects as a whole, 91.7\% of the MPLCs involve source files written in two PLs. 
  \item The change complexity (in terms of the number of lines of code, source files, directories modified, and Entropy) of MPLCs is significantly higher than that of non-MPLCs in all selected projects. 
  \item In 80\% of the selected projects, the issues fixed in MPLCs take longer (by 8.0\% to 124.7\%) to be resolved than the issues fixed in non-MPLCs;
  \item In 80\% of the selected projects, the source files that have been modified in MPLCs are more bug-prone (by 17.2\% to 3681.8\%) in terms of defect density than source files that have never been modified in MPLCs.
\end{itemize}

Based on the results of this empirical study, our future research will focus on the following directions: First, one promising direction is to investigate how MPLCs and software architecture interplay in MPL software systems. For instance, we may look into whether MPLCs are related to architectural technical debt \cite{LiLiAv2015,liAvLi2015}. 
Second, we plan to use MPLCs as an additional factor to enhance bug prediction models for MPL systems based on existing models.
Finally, we will replicate this study by constructing a large-scale dataset, in which more projects written in diverse PLs will be included to balance the use of different PLs.
\balance

\bibliographystyle{ieeetr}

\bibliography{ICPC2021}

\begin{thebibliography}{10}

\bibitem{GrAbJa2020}
M.~Grichi, M.~Abidi, F.~Jaafar, E.~E. Eghan, and B.~Adams, ``On the impact of
  interlanguage dependencies in multilanguage systems empirical case study on
  java native interface applications {(JNI)},'' {\em IEEE Transactions on
  Reliability}, pp.~1--13, 2020.
\newblock
  doi:\href{http://dx.doi.org/10.1109/TR.2020.3024873}{10.1109/TR.2020.3024873}.

\bibitem{RaPoFiDe2014}
B.~Ray, D.~Posnett, V.~Filkov, and P.~Devanbu, ``A large scale study of
  programming languages and code quality in github,'' in {\em Proceedings of
  the 22nd ACM SIGSOFT International Symposium on Foundations of Software
  Engineering (FSE'14)}, pp.~155--165, ACM, 2014.

\bibitem{MaBa2015}
P.~Mayer and A.~Bauer, ``An empirical analysis of the utilization of multiple
  programming languages in open source projects,'' in {\em Proceedings of the
  19th International Conference on Evaluation and Assessment in Software
  Engineering (EASE'15)}, p.~article 4, ACM, 2015.

\bibitem{KoWiLo2016}
P.~S. Kochhar, D.~Wijedasa, and D.~Lo, ``A large scale study of multiple
  programming languages and code quality,'' in {\em Proceedings of the IEEE
  23rd International Conference on Software Analysis, Evolution, and
  Reengineering (SANER'16)}, pp.~563--573, 2016.

\bibitem{Ma2017}
P.~Mayer, ``A taxonomy of cross-language linking mechanisms in open source
  frameworks,'' {\em Computing}, vol.~99, no.~7, pp.~701--724, 2017.

\bibitem{MaKiLe2017}
P.~Mayer, M.~Kirsch, and M.~A. Le, ``On multi-language software development,
  cross-language links and accompanying tools: a survey of professional
  software developers,'' {\em Journal of Software Engineering Research and
  Development}, vol.~5, no.~1, pp.~1--33, 2017.

\bibitem{AbGrKh2019}
M.~Abidi, M.~Grichi, and F.~Khomh, ``Behind the scenes: developers' perception
  of multi-language practices,'' in {\em Proceedings of the 29th Annual
  International Conference on Computer Science and Software Engineering
  (CASCON'19)}, pp.~72--81, IBM Corp., 2019.

\bibitem{KoLiWo2006}
K.~Kontogiannis, P.~Linos, and K.~Wong, ``Comprehension and maintenance of
  large-scale multi-language software applications,'' in {\em Proceedings of
  the 22nd IEEE International Conference on Software Maintenance (ICSM'06)},
  pp.~497--500, IEEE, 2006.

\bibitem{AbRaOpKh2021}
M.~Abidi, M.~S. Rahman, M.~Openja, and F.~Khomh, ``Are multi-language design
  smells fault-prone? an empirical study,'' {\em ACM Transactions on Software
  Engineering and Methodology}, vol.~30, no.~3, p.~Article No. 29, 2021.

\bibitem{ShMiAb2019}
A.~Shatnawi, H.~Mili, M.~Abdellatif, Y.-G. Guéhéneuc, N.~Moha, G.~Hecht,
  G.~E. Boussaidi, and J.~Privat, ``Static code analysis of multilanguage
  software systems,'' tech. rep., 2019.

\bibitem{kaIsIz2019}
M.~Kargar, A.~Isazadeh, and H.~Izadkhah, ``Multi-programming language software
  systems modularization,'' {\em Computers \& Electrical Engineering}, vol.~80,
  p.~106500, 2019.

\bibitem{GrEgAd2020}
M.~Grichi, E.~E. Eghan, and B.~Adams, ``On the impact of multi-language
  development in machine learning frameworks,'' in {\em Proceedings of the 36th
  IEEE International Conference on Software Maintenance and Evolution
  (ICSME'20)}, pp.~546--556, IEEE, 2020.

\bibitem{BeHoMaViVi2019}
E.~D. Berger, C.~Hollenbeck, P.~Maj, O.~Vitek, and J.~Vitek, ``On the impact of
  programming languages on code quality: A reproduction study,'' {\em ACM
  Transactions on Programming Languages and Systems}, vol.~41, no.~4,
  p.~Article 21, 2019.

\bibitem{BhNe2011}
P.~Bhattacharya and I.~Neamtiu, ``Assessing programming language impact on
  development and maintenance: a study on {C} and {C}++,'' in {\em Proceedings
  of the 33rd International Conference on Software Engineering (ICSE'11)},
  pp.~171--180, 2011.

\bibitem{AbKhGu2019a}
M.~Abidi, F.~Khomh, and Y.-G. Guéhéneuc, ``Anti-patterns for multi-language
  systems,'' in {\em Proceedings of the 24th European Conference on Pattern
  Languages of Programs (EuroPLoP'19)}, p.~Article 42, ACM, 2019.

\bibitem{AbKhGu2019b}
M.~Abidi, M.~Grichi, F.~Khomh, and Y.-G. Guéhéneuc, ``Code smells for
  multi-language systems,'' in {\em Proceedings of the 24th European Conference
  on Pattern Languages of Programs (EuroPLoP'19)}, p.~Article 12, ACM, 2019.

\bibitem{kaIsIz2020}
M.~Kargar, A.~Isazadeh, and H.~Izadkhah, ``Improving the modularization quality
  of heterogeneous multi-programming software systems by unifying structural
  and semantic concepts,'' {\em Journal of Supercomputing}, vol.~76, no.~1,
  pp.~87--121, 2020.

\bibitem{RuHo2009}
P.~Runeson and M.~H{\"o}st, ``Guidelines for conducting and reporting case
  study research in software engineering,'' {\em Empirical Software
  Engineering}, vol.~14, no.~2, pp.~131--164, 2009.

\bibitem{Ba1992}
V.~R. Basili, ``Software modeling and measurement: The goal/question/metric
  paradigm,'' tech. rep., 1992.

\bibitem{Ha2009}
A.~E. Hassan, ``Predicting faults using the complexity of code changes,'' in
  {\em Proceedings of the 31st International Conference on Software Engineering
  (ICSE'09)}, pp.~78--88, IEEE, 2009.

\bibitem{LiLiLiMoLi2020}
Z.~Li, P.~Liang, D.~Li, R.~Mo, and B.~Li, ``Is bug severity in line with bug
  fixing change complexity?,'' {\em International Journal of Software
  Engineering and Knowledge Engineering}, vol.~30, no.~11\&12, pp.~1779--1800,
  2020.

\bibitem{Fi2013}
A.~Field, {\em Discovering Statistics using IBM SPSS Statistics}.
\newblock Singapore: Sage Publications Ltd., fourth~ed., 2013.

\bibitem{LiLiLi2017}
Z.~Li, P.~Liang, and B.~Li, ``Relating alternate modifications to defect
  density in software development,'' in {\em Proceedings of the 39th
  International Conference on Software Engineering Companion (ICSE'17)},
  pp.~308--310, IEEE, 2017.

\bibitem{MoCaKaXiFe2016}
R.~Mo, Y.~Cai, R.~Kazman, L.~Xiao, and Q.~Feng, ``Decoupling level: A new
  metric for architectural maintenance complexity,'' in {\em Proceedings of the
  38th International Conference on Software Engineering (ICSE'16)},
  pp.~499--510, ACM, 2016.

\bibitem{WoCaKiDa2011}
S.~Wong, Y.~Cai, M.~Kim, and M.~Dalton, ``Detecting software modularity
  violations,'' in {\em Proceedings of the 33rd International Conference on
  Software Engineering (ICSE'11)}, pp.~411--420, IEEE, 2011.

\bibitem{LiLiAv2015}
Z.~Li, P.~Liang, and P.~Avgeriou, ``Architectural technical debt identification
  based on architecture decisions and change scenarios,'' in {\em Proceedings
  of the 12th Working IEEE/IFIP Conference on Software Architecture
  (WICSA'15)}, pp.~65--74, IEEE, 2015.

\bibitem{liAvLi2015}
Z.~Li, P.~Avgeriou, and P.~Liang, ``A systematic mapping study on technical
  debt and its management,'' {\em Journal of Systems and Software}, vol.~101,
  no.~3, pp.~193--220, 2015.

\end{thebibliography}


\end{document}